\documentclass[runningheads]{llncs}
\usepackage[T1]{fontenc}
\usepackage{graphicx,verbatim}
\usepackage{amsmath,amssymb,amsfonts}
\usepackage{algorithmic}
\usepackage{graphicx}
\usepackage{textcomp}
\usepackage{subfigure}
\usepackage[inkscapelatex=false]{svg}
\usepackage{mathrsfs}

\DeclareMathOperator*{\argmax}{arg\,max}

\DeclareMathOperator{\realset}{\mathbb{R}}

\usepackage{booktabs}
\usepackage{bm}
\usepackage{mathtools}
\usepackage{dsfont}
\usepackage{algorithm}
\usepackage{algorithmic}
\usepackage{multirow}

\usepackage{enumitem}
\usepackage{xcolor}
\usepackage{color, colortbl}
\definecolor{Gray}{gray}{0.9}
\usepackage{pifont}

\usepackage{hyperref}
\usepackage[capitalize]{cleveref}
\crefname{section}{Sec.}{Secs.}
\Crefname{section}{Section}{Sections}
\Crefname{table}{Table}{Tables}
\crefname{table}{Table.}{Tables.}

\crefformat{equation}{#2(#1)#3}
\Crefformat{equation}{#2(#1)#3}
\crefmultiformat{equation}{#2(#1)#3}{ and #2(#2)#3}{, #2(#2)#3}{ and #2(#2)#3}
\Crefmultiformat{equation}{#2(#1)#3}{ and #2(#2)#3}{, #2(#2)#3}{ and #2(#2)#3}
\crefrangeformat{equation}{(#3#1#4) to (#5#2#6)}

\DeclarePairedDelimiterX{\infdivx}[2]{(}{)}{%
  #1\;\delimsize\|\;#2%
}

\newcommand{\Expect}[2]{\mathop{\mathbb{E}} \limits_{#1}\left[ #2 \right]}
\newcommand{\cceg}{\mathrm{\Delta}_\varepsilon}
\newcommand{\predconfv}{\hat{\mathbf{p}}_\mathcal{Y}(\mathbf{D},\mathbf{x},a)}
\newcommand{\popconfv}{\Bar{\mathbf{p}}_\mathcal{Y}(\mathbf{D},\mathbf{x},a)}
\newcommand{\subgconfv}{\Tilde{\mathbf{p}}_\mathcal{Y}(\mathbf{D},\mathbf{x},a)}
\newcommand{\mergeconfv}{\Check{\mathbf{p}}_\mathcal{Y}(\mathbf{D},\mathbf{x},a)}
\newcommand{\prednullnullconfv}{\hat{\mathbf{p}}_\mathcal{Y}(\mathbf{D},\cdot,\cdot)}
\newcommand{\prednullconfv}{\hat{\mathbf{p}}_\mathcal{Y}(\mathbf{D},\cdot,a)}
\newcommand{\ivec}{\mathbf{i}}

\begin{document}
\title{Exposing and Mitigating Calibration Biases and Demographic Unfairness in MLLM Few-Shot In-Context Learning for Medical Image Classification}
\titlerunning{Exposing and Mitigating Calibration Biases in MLLM In-Context Learning}
\author{
Xing Shen\inst{1,2} \and
Justin Szeto\inst{1,2} \and
Mingyang Li\inst{3} \and
Hengguan Huang\inst{4} \and \\
Tal Arbel\inst{1,2}
}
\authorrunning{X. Shen et al.}
%
\institute{
Centre for Intelligent Machines, McGill University, Montreal, Canada \\ \email{xing.shen@mail.mcgill.ca, tal.arbel@mcgill.ca} \and
Mila -- Quebec AI Institute, Montreal, Canada \and 
Stanford University, Stanford, USA \and 
University of Copenhagen, Copenhagen, Denmark\\
\email{hengguan.huang@sund.ku.dk}
}
    
\maketitle              
\begin{abstract}
Multimodal large language models (MLLMs) have enormous potential to perform few-shot in-context learning in the context of medical image analysis. However, safe deployment of these models into real-world clinical practice requires an in-depth analysis of the accuracies of their predictions, and their associated calibration errors, particularly across different demographic subgroups. In this work, we present the first investigation into the calibration biases and demographic unfairness of MLLMs' predictions and confidence scores in few-shot in-context learning for medical image classification. We introduce \textbf{CALIN}, an inference-time calibration method designed to mitigate the associated biases. Specifically, CALIN estimates the amount of calibration needed, represented by calibration matrices, using a bi-level procedure: progressing from the population level to the subgroup level prior to inference. It then applies this estimation to calibrate the predicted confidence scores during inference. Experimental results on three medical imaging datasets: PAPILA for fundus image classification, HAM10000 for skin cancer classification, and MIMIC-CXR for chest X-ray classification demonstrate CALIN's effectiveness at ensuring fair confidence calibration in its prediction, while improving its overall prediction accuracies and exhibiting minimum fairness-utility trade-off. Our codebase can be found at \url{https://github.com/xingbpshen/medical-calibration-fairness-mllm}.


\end{abstract}
\section{Introduction}
Image-text to text foundation models, particularly multimodal large language models (MLLMs, or referred to as large multimodal models, LMMs), such as OpenAI GPT-4o and Google Gemini~\cite{hurst2024gpt,team2024gemini}, have demonstrated strong generalization capabilities and achieved state-of-the-art performance across numerous tasks. Furthermore, advances in few-shot in-context learning (FS-ICL) enables MLLMs to solve new tasks by simply being \textit{prompted} with a few examples of question-answer pairs \cite{baldassini2024makes,wang2023seggpt,brown2020language}. The success of MLLMs and FS-ICL methods has led to applications in medical imaging contexts, including cancer pathology classification \cite{ferber2024context}, where they have shown promising results while reducing or eliminating the need for the extensive training or fine-tuning typically required by traditional deep learning methods. However, in the context of medical imaging, ensuring debiased and fair machine learning models, particularly with respect to both prediction utility and confidence calibration across different demographic subgroups, is essential in order to safely deploy these models in real clinical contexts \cite{zong2023medfair,jin2024fairmedfm,shui2023mitigating}. The associated risks include trusting prediction uncertainties that can potentially indicate high confidence in wrong assertions, for example, or presenting disparities in model performance across groups which can lead to potential harm to underrepresented groups. Despite these risks, investigations into calibration biases in MLLMs under FS-ICL setting, as well as strategies to accurately overcome their errors and biases in medical imaging, remains unexplored. This limits their practical use and reliability in real-world clinical settings. 

Enforcing calibration fairness under FS-ICL setting poses unique methodological challenges. The lack of an additional training/validation set with an adequate amount of labeled data for different subgroups renders widely adopted optimization-based calibration methods impractical \cite{kull2019beyond,guo2017calibration}. In addition, the most powerful state-of-the-art MLLMs are typically large-scale black-box models (e.g., GPT-4o, Gemini 1.5, Claude 3.5 Sonnet), making debiasing methods requiring additional access of their internal parameters infeasible~\cite{he2024prompt}.

\begin{figure}[ht!]
    \centerline{\includegraphics[width=0.9\linewidth]{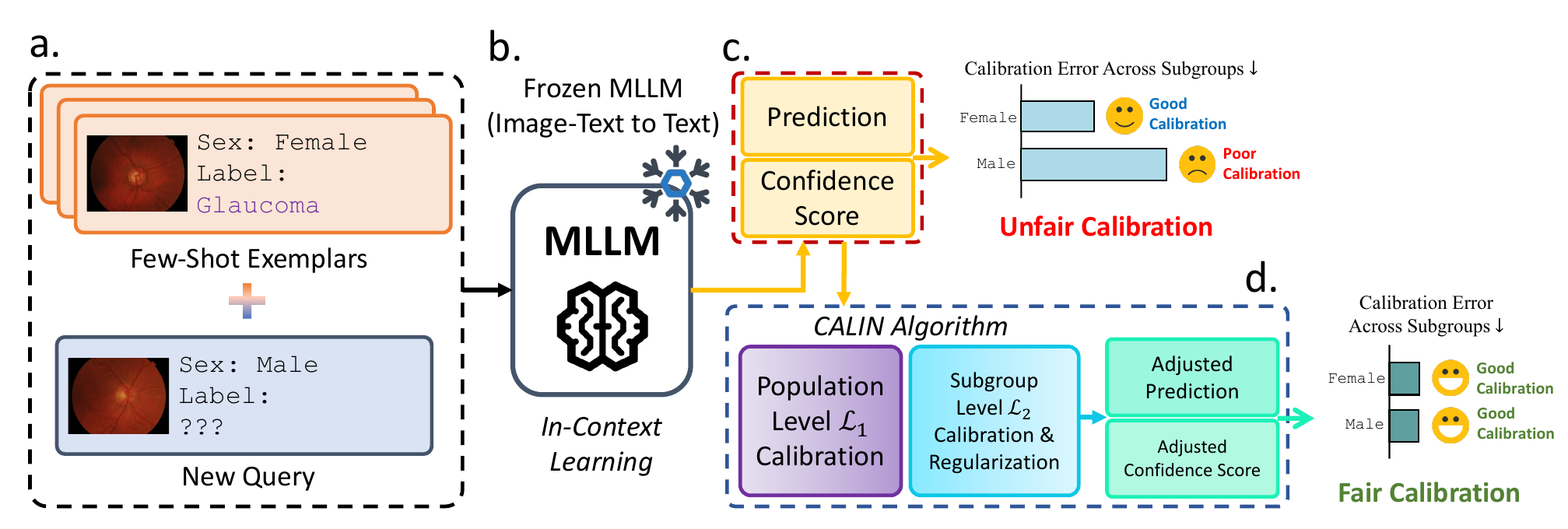}}
    \caption{Overview of CALIN for medical image classification with FS-ICL: (a) The MLLM takes as input a set of few-shot exemplars, each comprising an image, an associated attribute, and a label, along with a new query image and its attribute for label prediction. (b) MLLM predicts the label for the query image and the associated confidence score is calculated. (c) The predictions from the MLLM exhibit confidence calibration biases, leading to demographic disparities. (d) CALIN adjusts the confidence scores to mitigate calibration errors and improves fairness across demographic groups.}
    \label{fig:fs_icl}
\end{figure}

In order to fill the gap and enable the trustworthy deployment of FS-ICL methods, this work investigates the calibration unfairness of MLLM under FS-ICL, exposing their biases and limitations in the context of medical image classification. To address current challenges, we propose \textbf{CALIN}, a novel training-free algorithm that automatically calibrates MLLM's predictions and their associated confidence scores, and enforces fairness across demographic subgroups at inference. CALIN (see \cref{fig:fs_icl}) uses a \textit{bi-level} procedure: progressing from the \textit{population level} to the \textit{subgroup level}, ensuring an accurate and stable adjustment estimation procedure for fair calibration across subgroups. Extensive experiments are performed on three publicly available medical imaging datasets--PAPILA \cite{kovalyk2022papila} for fundus image classification, HAM10000 \cite{tschandl2018ham10000} for skin cancer classification, and MIMIC-CXR \cite{johnson2019mimicjpg} for chest X-ray classification. Experimental results expose calibration biases in the MLLM under FS-ICL, and validate CALIN's effectiveness at: (i) mitigating the calibration gap between demographic subgroups, (ii) providing more reliable confidence scores over the entire population, (iii) improving prediction accuracies, and (iv) exhibiting a minimum fairness-utility trade-off. Detailed ablation studies further validate the necessity of the bi-level method in producing reliable and fair confidence calibrations.

\section{Background on FS-ICL and Calibration Biases}
We formally define the few-shot in-context learning (FS-ICL) setting and demographic calibration biases. At inference, a few-shot exemplar dataset with $N$ samples (e.g., $N\leq5$) is presented, represented as 3-tuples $\mathcal{D_\mathrm{fs}}:=\{(X_i, A_i,Y_i)\}_{i=1}^N$, where $X_i$ is a random variable representing the medical image of the patient, $A_i$ is a random variable representing the sensitive attribute (e.g., sex, age) of that patient, $Y_i$ is a random variable representing the label of the image. Every tuple in $\mathcal{D_\mathrm{fs}}$ follows the same task distribution denoted as $(X_i,A_i,Y_i)\sim P_\tau(X,A,Y)$. Given a new query $(X,A)\sim P_\tau(X,A)$ and a predictive model $f(\cdot)$ with fixed parameters, the new prediction $\hat{Y}$ from few-shot in-context learning is $\hat{Y}=f(\mathcal{D}_\mathrm{fs},X,A).$

The demographic calibration bias can be defined as the confidence calibration error gap (CCEG, $\cceg$) between subgroups under a sensitive attribute \cite{jin2024fairmedfm}. Formally, for a given demographic attribute $A$, the gap $\cceg$ can be expressed as:
\begin{gather}
    \varepsilon(a) = \mathbb{E}\left[\left|\Pr[Y = \hat{Y} \mid \hat{p}, A = a] - \hat{p}\right|\right], \label{eq:cceg_1}\\
    \cceg(A) = \Expect{(a, b) \sim \mathcal{U}\left(\{(a, b) \mid a, b \in \mathcal{A}, a \neq b\}\right)}{|\varepsilon(a) - \varepsilon(b)|}, \label{eq:cceg_2}
\end{gather}  
where $\hat{p}$ is the predicted confidence for the prediction $\hat{Y}$, $Y$ is the ground-truth label, $\mathcal{A}=\mathrm{Val}(A)$ is the support of $A$, and $(a,b)$ is a 2-tuple of values sampled uniformly from $\mathcal{A} \times \mathcal{A}$ such that $a \neq b$. A perfectly fair model has $\cceg(A)=0$.

\section{\textbf{CALIN}: Intergroup \underline{C}onfidence \underline{Ali}gnment From \underline{N}ull-Input Calibration}
To overcome calibration errors and biases under the FS-ICL setting and to ensure calibration fairness among subgroups, we propose \textbf{CALIN}, an \textit{inference-time calibration} method that contains a bi-level procedure -- from \textit{population-level} to \textit{subgroup-level}. The goal is to provide fair and reliable confidence without requiring an additional training/validation set or access to the MLLM's parameters.

\subsection{Notations}
We assume that the predictive model is implemented by a pretrained frozen multimodal large language model $f_\mathrm{MLLM}(\cdot)$ (e.g., GPT-4o and Gemini-1.5 \cite{hurst2024gpt,team2024gemini}) that takes as input a set of \textit{multimodal prompts} (image and text). We define a template $\varphi$ that has fields for an image $X$, attributes $A$, and the label $Y$, though some may be left empty, to generate multimodal prompts. For example, $\varphi(X=\mathbf{x},A=\mathrm{Male},Y=\mathrm{Negative})$ is mapped to: ``Does the fundus $\mathbf{x}$ of a male show glaucoma? Negative'' (see  \cref{tab:prompts} for more examples). During inference, the model is provided with the multimodal prompt for the new query $\varphi(X=\mathbf{x},A=a,\cdot)$ along with  FS-ICL (few-shot) exemplars $\mathbf{D}:=\{(X_i=\mathbf{x}_i,A_i=a_i,Y_i=y_i)|(X_i,A_i,Y_i)\in\mathcal{D}_\mathrm{fs}\}$. The MLLM's predicted probability for $\hat{Y}$ being $y$ given the inputs is denoted $\hat{p}_y(\mathbf{D},\mathbf{x},a)$ and estimated as follows:
\begin{align}
    \underbrace{\Pr\left[\hat{Y}=y\mid \mathbf{D},X=\mathbf{x},A=a\right]}_{\hat{p}_y(\mathbf{D},\mathbf{x},a)}= \frac{\Pr\left[\hat{T}=y \mid \mathbf{D},X=\mathbf{x},A=a \right]}{\sum_{y_j\in\mathcal{Y}}\Pr\left[\hat{T}=y_j \mid \mathbf{D},X=\mathbf{x},A=a \right]}. \label{eq:renorm}
\end{align}
Here, $\hat{T}=f_\mathrm{MLLM}\big( \{\varphi(X_i,A_i,Y_i)|(X_i,A_i,Y_i)\in\mathcal{D}_\mathrm{fs}\}\cup\{\varphi(X,A, \cdot)\} \big)$ is a random variable denoting the predicted next-token, and $\mathcal{Y}=\mathrm{Val}(Y)$. We additionally define a vector $\predconfv\in\realset^{|\mathcal{Y}|}$, where each dimension represents the probability of the prediction belonging to a specific class $y_j \in \mathcal{Y}$.

\begin{table}[h]
\centering
\fontsize{8pt}{9.6pt}\selectfont
\caption{Multimodal prompts $\varphi$ under different inputs for fundus image classification. The left illustrates a datapoint containing the fundus image $X=\mathbf{x}$, the value of the attribute $A=\mathrm{Male}$, and the label $Y=\mathrm{Negative}$. The right illustrates the corresponding prompts. For cases $\varphi(\cdot,A,\cdot)$ and $\varphi(\cdot,\cdot,\cdot)$, we do not input the image to the MLLM.}
\label{tab:prompts}
\begin{tabular}{cll}
\hline
\rowcolor{Gray} \textbf{Example Prompts} $\varphi$ & & \\ \hline
\multicolumn{1}{c|}{\multirow{3}{*}{\includegraphics[width=1.3cm, height=1cm]{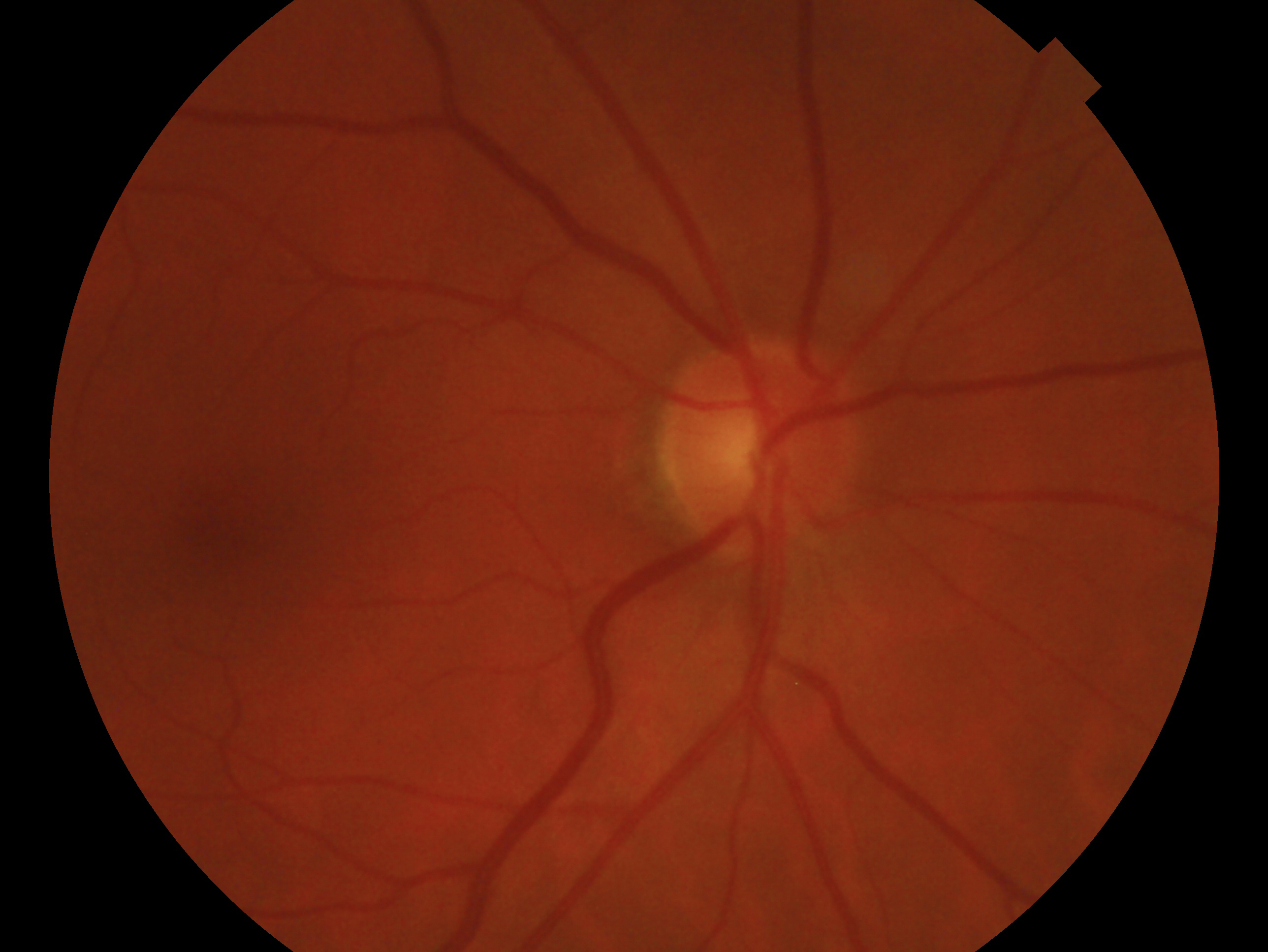}}} & $\varphi(X,A,Y)$ & \textit{Does the fundus of a \textbf{male} show glaucoma? \textbf{Negative}} \\
\multicolumn{1}{c|}{}                     & $\varphi(X,A,\cdot)$ & \textit{Does the fundus of a \textbf{male} show glaucoma?} \\
\multicolumn{1}{c|}{}                     & $\varphi(\cdot,A,\cdot)$ & \textit{Does an arbitrary fundus of a \textbf{male} show glaucoma?} \\
\multicolumn{1}{c|}{\textbf{Male} with \textbf{no glaucoma}}                & $\varphi(\cdot,\cdot,\cdot)$ & \textit{Does an arbitrary fundus show glaucoma?} \\ \hline
\end{tabular}
\end{table}

\subsection{Bi-Level Confidence Calibration}
The bi-level procedure used by CALIN can be intuitively thought of as first inferring the ``amount of calibration'' needed for the entire population (\textit{population-level}), then inferring the ``coarse'' amount of calibration needed for each subgroup (\textit{subgroup-level}). Information flows from the upper \textit{population-level} to regularize the lower \textit{subgroup-level} to provide an accurate and fair confidence calibration.

\paragraph{\textbf{Population-Level Calibration $\mathscr{L}_1$.}} Inspired by the findings of language model's prediction bias presented in \cite{zhao2021calibrate,han2023prototypical}, CALIN first infers the ``amount of calibration'' for the entire population to avoid prediction bias under FS-ICL. In this work, the amount of population-level calibration is defined by a diagonal matrix $\mathbf{U}\in\realset^{|\mathcal{Y}|\times |\mathcal{Y}|}$ (we call it \textit{calibration matrix} in this work), then the softmaxed linear transformation of $\predconfv$, determined by $\mathbf{U}$, is the $\mathscr{L}_1$ post-calibration confidence, given by $\popconfv=\mathrm{softmax}\left(\mathbf{U}\predconfv\right)$.

To determine $\mathbf{U}$ without the need of extra training/validation set, CALIN adopts a \textit{multimodal null-input probing} technique. Specifically, we ensure that the predicted confidence $\predconfv$ is aligned with a uniform distribution when a null (or ``content-free'', ``semantic-free'' \cite{zhao2021calibrate,ma2024fairness}) query $\varphi(\cdot,\cdot,\cdot)$ is fed into the MLLM. For a concrete binary classification example in \cref{tab:prompts}, when we neither provide the fundus image nor specify the sex of the patient, the MLLM's predicted confidence distribution should be uniform\footnote{We assume that it is impossible to identify the ground-truth label without observing the medical image $\mathbf{x}$.}. To this end, $\mathbf{U}$ is calculated based on the observed predicted confidence $\prednullnullconfv$ by the MLLM when we send null query $\varphi(\cdot,\cdot,\cdot)$ to it, given by $\mathbf{U}=\left(\mathrm{diag}\left( \prednullnullconfv \right)\right)^{-1}$.

\paragraph{\textbf{Subgroup-Level Calibration $\mathscr{L}_2$.}} 
$\mathscr{L}_1$ improves confidence calibration over the entire population. To capture the potential variations across subgroups, we propose \textit{subgroup-wise multimodal null-input probing} which aims to infer a set of calibration matrices $S:=\{\mathbf{S}_a|a\in\mathcal{A}\}$ for $\mathscr{L}_2$ calibration. Each matrix in $S$ focuses on calibrating one specific subgroup with sensitive attribute $A=a$. Borrowing from the intuition of multimodal null-input probing, subgroup-wise multimodal null-input probing finds $S$ such that the predicted confidence given an attribute-conditioned null query $\varphi(\cdot,A=a,\cdot)$ is uniform for all subgroups. Specifically, we calculate them based on the observed predicted confidence $\prednullconfv$ by the MLLM, given by $\mathbf{S}_a=\left(\mathrm{diag}\left( \prednullconfv \right)\right)^{-1}$ for all $a\in\mathcal{A}$. Then, $\subgconfv=\mathrm{softmax}\left( \mathbf{S}_a \predconfv \right)$ is the $\mathscr{L}_2$ post-calibration confidence for any new query.

\paragraph{\textbf{Regularizing $\mathscr{L}_2$ with $\mathscr{L}_1$.}}
While $\mathscr{L}_2$ calibration aims to achieve subgroup level confidence alignment, relying solely on $\mathscr{L}_2$ may not guarantee accurate calibration. This is because the language model's inherent prompt bias \cite{xu2024take,cao2021knowledgeable} can lead to inaccurate and unstable estimation of calibration matrices, particularly since the $\mathscr{L}_2$'s probing prompt $\varphi(\cdot,A,\cdot)$ includes additional semantic information by conditioning on sensitive attributes. To mitigate this issue, we leverage $\mathscr{L}_1$ as a regularization mechanism, allowing the final calibration to capture subgroup variability and also penalizing anomalies. Specifically, we calculate a new set of calibration matrices $C:=\{\mathbf{C}_a|a\in\mathcal{A}\}$ using exponential decay: When the estimated $\mathscr{L}_2$ calibration $\mathbf{S}_a$ extremely diverges (due to unstable estimation) from $\mathscr{L}_1$ calibration $\mathbf{U}$, the final calibration will be more aligned with $\mathscr{L}_1$, otherwise, the final calibration will be more aligned with $\mathscr{L}_2$. The decay rate is governed by $(\sqrt{\alpha+1})^{-1}$ where $\alpha$ is the maximum observed deviation across subgroups, calculated by $\alpha=\max_a\{\| \mathbf{S}_a\ivec- \mathbf{U}\ivec \|_\infty\}$ where $\|\cdot\|_\infty$ denotes the infinity-norm. The final calibration matrices are given by:
\begin{gather}
    \mathbf{c}_a=\mathbf{U}\mathbf{i} +\left( \mathbf{S}_a\ivec- \mathbf{U}\ivec \right)\odot \exp\left({-(\sqrt{\alpha+1})^{-1} \cdot |\mathbf{S}_a\ivec- \mathbf{U}\ivec|}\right), \label{eq:C_a_1} \\
    \mathbf{C}_a=\mathrm{diag}(\mathbf{c}_a), \quad \forall a\in\mathcal{A}, \label{eq:C_a_2}
\end{gather}
where $\ivec=\mathbf{1}_{|\mathcal{Y}|}$ is a vector with $|\mathcal{Y}|$ ones, $\odot$ denotes the element-wise product. We obtain the post-calibration confidence $\mergeconfv=\mathrm{softmax}\left(\mathbf{C}_a\predconfv\right)$. Given the denoted vector construction $\mergeconfv=[\Check{p}_{y_j}(\mathbf{D},\mathbf{x},a)|y_j\in\mathcal{Y}]$ we can get the adjusted predicted label $\Check{y}=\argmax_{y_j\in\mathcal{Y}}\left\{\Check{p}_{y_j}(\mathbf{D},\mathbf{x},a) \right\}$. The algorithm of CALIN is shown in \cref{alg:calin}.

\begin{algorithm}[]
\fontsize{9pt}{10pt}\selectfont
\caption{CALIN for Fair Confidence Calibration Under FS-ICL}
\label{alg:calin}
\begin{algorithmic}[1]
    \REQUIRE Few-shot $\mathbf{D}$, model $f_\mathrm{MLLM}$, prompt template $\varphi$, demographic values $\mathcal{A}$
    \ENSURE Calibration matrices $C$
    \STATE Compute $\prednullnullconfv$ using \cref{eq:renorm} with $f_\mathrm{MLLM}$, $\mathbf{D}$, $\varphi(\cdot,\cdot,\cdot)$
    \STATE Compute $\mathbf{U}=\left(\mathrm{diag}\left( \prednullnullconfv \right)\right)^{-1}$ \quad\quad\quad\quad\quad\quad\quad\quad\quad\quad\; \#\textit{Population-Level}\#
    \FOR{\( a \) in $\mathcal{A}$}
        \STATE Compute $\prednullconfv$ using \cref{eq:renorm} with $f_\mathrm{MLLM}$, $\mathbf{D}$, $\varphi(\cdot,A=a,\cdot)$
        \STATE Compute $\mathbf{S}_a=\left(\mathrm{diag}\left( \prednullconfv \right)\right)^{-1}$\quad\quad\quad\quad\quad\quad\quad\quad\quad\quad\. \#\textit{Subgroup-Level}\#
    \ENDFOR
    \STATE Compute $\mathbf{C}_a$ using \cref{eq:C_a_1} and \cref{eq:C_a_2} with $\mathbf{U}$ and $\mathbf{S}_a$, add $\mathbf{C}_a$ to $C$. For all $a\in\mathcal{A}$
    \RETURN $C$
    \REQUIRE Few-shot $\mathbf{D}$, new query medical image $\mathbf{x}^*$, demographic value $a^*\in\mathcal{A}$, model $f_\mathrm{MLLM}$, prompt template $\varphi$, calibration matrix $\mathbf{C}_{a^*}\in C$
    \ENSURE Adjusted prediction \(\Check{y}\) and its calibrated confidence $\Check{p}$
    \STATE Compute $\hat{\mathbf{p}}_\mathcal{Y}(\mathbf{D},\mathbf{x}^*, a^*)$ using \cref{eq:renorm} with $f_\mathrm{MLLM}$, $\mathbf{D}$, $\varphi(X=\mathbf{x}^*,A=a^*,\cdot)$
    \STATE Compute $\Check{\mathbf{p}}_\mathcal{Y}(\mathbf{D},\mathbf{x}^*,a^*)=\mathrm{softmax}\left( \mathbf{C}_{a^*}\hat{\mathbf{p}}_\mathcal{Y}(\mathbf{D},\mathbf{x}^*, a^*) \right)$  \quad\quad\quad\#\textit{Inference-Time}\#
    \STATE Assign vector elements $[\Check{p}_{y_j}(\mathbf{D},\mathbf{x}^*,a^*)|y_j\in\mathcal{Y}]=\Check{\mathbf{p}}_\mathcal{Y}(\mathbf{D},\mathbf{x}^*,a^*)$
    \RETURN $\Check{y}=\argmax_{y_j\in\mathcal{Y}}\left\{\Check{p}_{y_j}(\mathbf{D},\mathbf{x}^*,a^*) \right\}$ and $\Check{p}=\Check{p}_{\Check{y}}(\mathbf{D},\mathbf{x}^*,a^*)$
\end{algorithmic}
\end{algorithm}

\section{Experiments and Results}
Experiments are designed to showcase the effectiveness of CALIN in mitigating confidence calibration bias in MLLM under FS-ICL on 3 medical imaging datasets: (i) PAPILA \cite{kovalyk2022papila}, (ii) HAM10000 \cite{tschandl2018ham10000}, (iii) MIMIC-CXR \cite{johnson2019mimicjpg}.

\paragraph{Datasets, Configuration and Implementation Details.} The \underline{PAPILA dataset} \cite{kovalyk2022papila} is a glaucoma classification dataset consisting of patient fundus images, along with their sex and ages. 364 patients are randomly chosen as the test set, including 118 male and 246 female patients, with 146 young (age $<$ 60) patients, and 218 elder (age $\geq$ 60) patients. The images are binary-labeled, indicating the diagnosis of glaucoma. The \underline{HAM10000 dataset} \cite{tschandl2018ham10000} is a large-scale skin lesion classification dataset, consisting of dermatoscopic images of pigmented skin lesions, along with the patients' sex and ages. 1,062 patients are randomly chosen for the test set, including 566 male and 496 female patients, with 472 young patients and 590 elder patients. A binary label indicates the diagnosis as malignant or benign. The \underline{MIMIC-CXR dataset} \cite{johnson2019mimicjpg} is large-scale dataset of chest radiographs with structured labels. 1,062 randomly chosen patients make up the test set, including 547 male and 488 female patients, with 439 young patients and 623 elder patients. A binary label indicates the diagnosis of pleural effusion. \underline{All experiments} are conducted using GPT-4o-mini. Human review of input data is disabled on Azure OpenAI Service to comply with PhysioNet's guidelines for responsible use of MIMIC-CXR with GPT \cite{mimic-gpt-responsible-use}. We treat both sex and age as sensitive attributes. In each context, 4 additional patients are randomly selected from the dataset, apart from the test set, to serve as few-shot exemplars.

\subsection{Main Results}
\begin{figure}[h]
  \centering
  \includegraphics[width=0.3\linewidth]{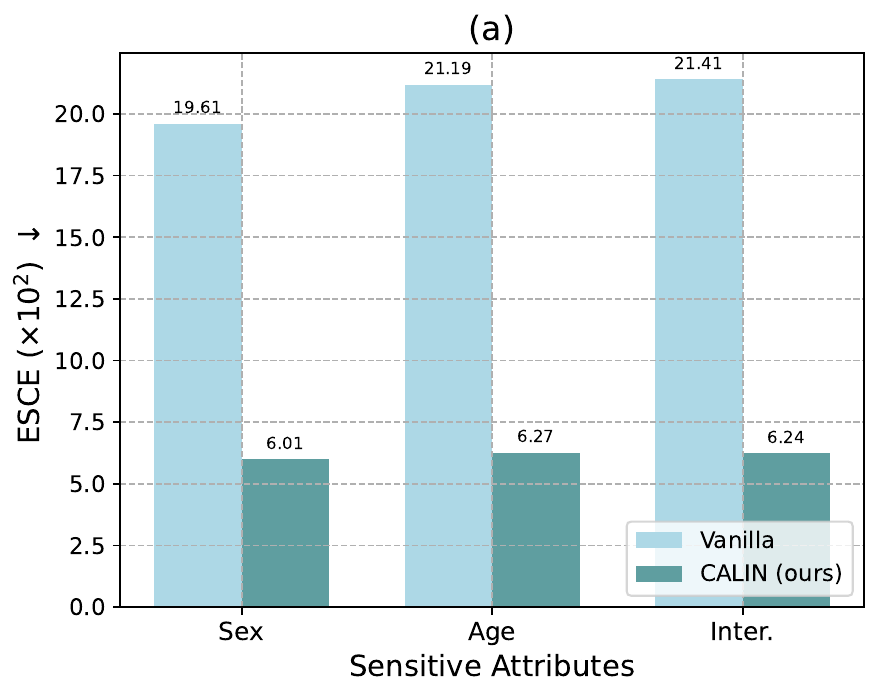}
  \includegraphics[width=0.293\linewidth]{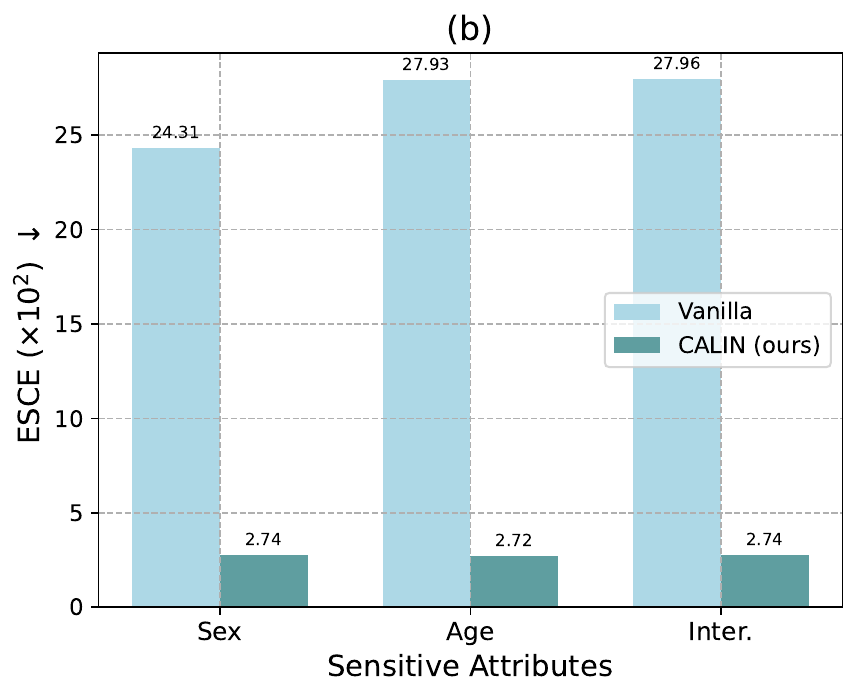}
  \includegraphics[width=0.293\linewidth]{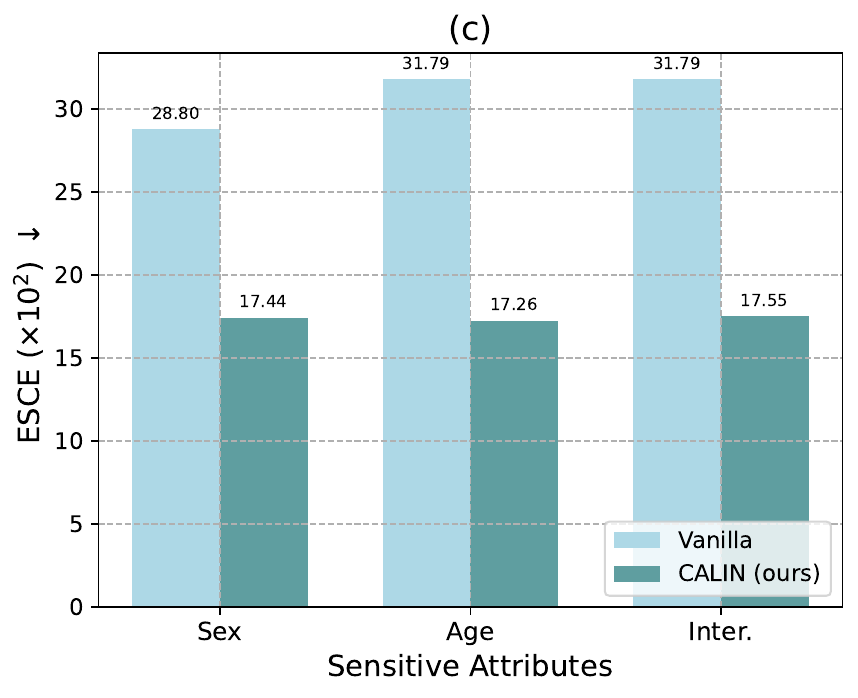}
  \caption{Results on equity-scaling measure of calibration error (ESCE) on 3 datasets: (a) PAPILA \cite{kovalyk2022papila}, (b) HAM10000 \cite{tschandl2018ham10000}, and (c) MIMIC-CXR \cite{johnson2019mimicjpg}. The proposed method consistently outperforms baseline method, vanilla FS-ICL, in terms of the fairness-utility trade-off.}
  \label{fig:esce}
\end{figure}

We consider 5 different metrics in the experiments: (i) classification accuracy ({\bf Acc.}), (ii) expected calibration error ({\bf ECE}) \cite{guo2017calibration} for quantifying the reliability of predicted confidence, (iii) mean equalized odd ratio \cite{mehrabi2021survey} between sex and age ($\overline{\mathbf{EOR}}$) for fairness evaluation, (iv) confidence calibration error gap ({\bf CCEG}, $\cceg$) for calibration fairness evaluation, and (v) equity-scaling measure \cite{luo2024harvard,tian2024fairseg,jin2024fairmedfm} of calibration error ({\bf ESCE}) for accessing the overall calibration performance adjusted by subgroups' performance. ECSE also quantifies the fairness-utility trade-off. Note that for CCEG, which is the main focus of this work, we consider fairness for the attributes of \textbf{sex}, \textbf{age}, and intersectional ({\bf Inter.}) fairness \cite{xu2024intersectional} of both attributes. Due to the lack of other valid baseline methods in this new problem setting, we compare the proposed method, CALIN, with the vanilla FS-ICL \cite{brown2020language}. Experiments on both methods were performed with the same exemplars and queries. Given GPT’s inherent stochasticity, and any future updates to GPT, might result in slight variability in the exact metric values. Experimental results in \cref{tab:main_results} and in \cref{fig:esce} indicate that CALIN consistently outperforms the vanilla method on all metrics (metric values are scaled by $\times 10^2$). Specifically, CALIN improves confidence calibration across the entire population, as indicated by a substantial reduction in ECE. More importantly, as evidenced by the notable decrease in CCEG across all datasets, CALIN effectively mitigates confidence calibration bias associated with demographic attributes. This is particularly evident for age and attribute intersection, where vanilla FS-ICL struggles with fairness issues across these demographic groups. In \cref{fig:esce}, CALIN demonstrates superior performance in the equity-scaling measure (ESCE), validating its minimal fairness-utility trade-off.

\begin{table}[h]
\centering
\fontsize{8pt}{9.6pt}\selectfont
\caption{Main results for 3 datasets. The proposed method consistently outperforms the vanilla FS-ICL \cite{brown2020language} baseline method according to all metrics, especially in terms of calibration across the population (ECE) and fair calibration across subgroups (CCEG).}
\tabcolsep=0.2cm
\label{tab:main_results}
\begin{tabular}{lcccccc}
\toprule
\multirow{2}{*}{Method} & \multirow{2}{*}{Acc. $\uparrow$} & \multirow{2}{*}{ECE $\downarrow$} & \multirow{2}{*}{$\overline{\mathrm{EOR}}$ $\uparrow$} & \multicolumn{3}{c}{CCEG $\cceg$ $\downarrow$} \\ \cmidrule{5-7} 
          &       &       &       & Sex  & Age   & Inter. \\ \midrule
\rowcolor{Gray}\textit{PAPILA} \cite{kovalyk2022papila}&       &       &       &      &       &        \\
Vanilla   & 78.30 & 19.13 & 20.00 & 4.84 & 19.37 & 15.15  \\
CALIN (proposed)& \textbf{78.57} & \textbf{5.97}  & \textbf{34.38} & \textbf{1.53} & \textbf{9.52}  & \textbf{6.14}   \\ \midrule
\rowcolor{Gray}\textit{HAM10000} \cite{tschandl2018ham10000}&       &       &       &      &       &        \\
Vanilla   & 74.76 & 23.70 & 70.51 & 5.01 & 30.25 & 20.66  \\
CALIN (proposed)& \textbf{74.76} & \textbf{2.68}  & \textbf{74.24} & \textbf{4.43} & \textbf{3.14}  & \textbf{3.11}   \\ \midrule
\rowcolor{Gray}\textit{MIMIC-CXR} \cite{johnson2019mimicjpg}&       &       &       &      &       &        \\
Vanilla   & 66.38 & 28.09 & 59.48 & 4.92 & 23.28 & 16.33  \\
CALIN (proposed)& \textbf{68.55} & \textbf{17.12} & \textbf{64.32} & \textbf{3.65} & \textbf{1.60} & \textbf{3.48} \\ \bottomrule
\end{tabular}
\end{table}

\begin{table}[htp!]
\centering
\fontsize{8pt}{9.6pt}\selectfont
\caption{Ablation study results on HAM10000. The bi-level approach outperforms baselines using single-level in most of the metrics.}
\tabcolsep=0.2cm
\label{tab:ablation}
\begin{tabular}{lcccccc}
\toprule
\multirow{2}{*}{Method} & \multirow{2}{*}{Acc. $\uparrow$} & \multirow{2}{*}{ECE $\downarrow$} & \multirow{2}{*}{$\overline{\mathrm{EOR}}$ $\uparrow$} & \multicolumn{3}{c}{CCEG $\cceg$ $\downarrow$} \\ \cmidrule{5-7} 
          &       &       &       & Sex  & Age   & Inter. \\ \midrule
$\mathscr{L}_1$ only    & 74.29 & 22.55 & 70.91 & 5.49 & 29.79 & 20.19  \\
$\mathscr{L}_2$ only & 64.88 & 14.13 & 72.66 & \textbf{0.83} & 22.73 & 16.43   \\ \midrule
Bi-level & \textbf{74.76} & \textbf{2.68}  & \textbf{74.24} & 4.43 & \textbf{3.14}  & \textbf{3.11} \\ \bottomrule
\end{tabular}
\end{table}

\subsection{CALIN Ablation Experiments}
Ablation experiments are chosen to validate the effectiveness of CALIN's bi-level framework, comparing its performance with baselines that use a single level ($\mathscr{L}_1$ or $\mathscr{L}_2$). Results in \cref{tab:ablation} illustrate that the \(\mathscr{L}_2\) baseline consistently outperforms \(\mathscr{L}_1\) in both fairness metrics, highlighting the importance of modeling subgroup variability. The bi-level framework further improves by a large margin across most metrics, demonstrating the effectiveness of regularizing $\mathscr{L}_2$ with $\mathscr{L}_1$.

\section{Limitations}
\paragraph{Model Limitation.} This study focuses on identifying and addressing calibration biases in modern multimodal large language models (MLLMs), specifically using the GPT family model in our experiments. Although our findings reveal the presence of such biases in this model, a comprehensive analysis across alternative MLLM architectures and varying model sizes remains an open direction for future research.

\paragraph{Task Limitation.} This study is restricted to medical image classification tasks where each label is represented by a single token. Such a formulation may be inadequate for tasks requiring multi-token labels. A possible workaround is to reformulate the task using single-token categorical options (e.g., \textit{A, B, C, D}). Additionally, a thorough investigation into the impact of different exemplar combinations is left to future work.

\section{Conclusion and Future Work}
In this paper, we examine MLLM's confidence calibration biases across demographic subgroups under FS-ICL, an area that remains unexplored in existing research. To address these biases, we introduce CALIN, a novel inference-time confidence calibration method. CALIN operates through a bi-level calibration procedure, effectively mitigating unfairness. Experimental results on three medical imaging datasets demonstrate that CALIN not only enhances fairness but also improves overall predictive performance and exhibits minimum fairness-utility trade-off. Future work should explore verbalized confidence \cite{xiong2024can}, and integrate prompt optimization \cite{lu2022fantastically,wu2024prompt} to refine in-context exemplars for each demographic subgroup for improved fairness.

\section{Ethics Statement}
This work investigates fairness and calibration in modern multimodal large language models when used as clinical decision support tools. To promote equitable and reliable outcomes, we propose a training-free mitigation approach that reduces potential biases in model predictions. All experiments involving health-related data were conducted in compliance with relevant ethical guidelines and regulatory standards. Necessary permissions were obtained to access and process the datasets used in this study, and care was taken to ensure that all data handling adhered to principles of privacy, security, and responsible AI use.

\begin{credits}
\subsubsection{\ackname} This work was supported in part by the Natural Sciences and Engineering Research Council of Canada, in part by the Canadian Institute for Advanced Research (CIFAR) Artificial Intelligence Chairs Program, in part by the Mila -- Quebec Artificial Intelligence Institute, in part by the Mila-Google Research Grant, and in part by the Canada First Research Excellence Fund, awarded to the Healthy Brains, Healthy Lives initiative at McGill University.
\end{credits}
%
%
%
\bibliographystyle{splncs04}
\bibliography{Paper}
\end{document}